\documentclass[apj]{emulateapj}

\usepackage{psfig}

\newcommand{\teff}{$T_{\mathrm{eff}}$}

\slugcomment{To appear in the Astrophysical Journal}

\shorttitle{Snapshot metallicity estimate of resolved stellar systems}
\shortauthors{Buzzoni et al.}

\begin{document}

\title{Snapshot metallicity estimate of resolved stellar systems\\
through Lick Fe5270 diagnostic}

\author{A. Buzzoni}
\affil{INAF - Osservatorio Astronomico di Bologna, Via Ranzani 1, 
40127 Bologna, Italy}

\author{E. Bertone and M. Chavez}
\affil{INAOE - Instituto Nacional de Astrof{\'\i}sica {\'O}ptica y Electr{\'o}nica, 
Luis Enrique Erro 1, 72840 Tonantzintla, Puebla, Mexico}
\email{alberto.buzzoni@oabo.inaf.it}

\begin{abstract}
We outline a new method to derive a ``snapshot'' metallicity estimate 
of stellar systems (providing one resolves at least the brightest part
of the CMD) just on the basis of low-resolution (i.e., 6--8~\AA\ FWHM) 
spectroscopy of a small stellar sample. Our method relies on the Fe5270 Lick index 
measurements and takes advantage of the special behavior of this spectral 
feature, that reaches its maximum strength among the ubiquitous component
of K-type giants. This makes the Fe5270$_{\rm max}$ estimate a robust and 
model-independent tracer of cluster [Fe/H], being particularly insensitive 
to the age of the stellar population.

A comparison of the Fe5270$_{\rm max}$ distribution derived from globular 
and open clusters, as well as from the field giant population in the Galaxy 
disk, confirms a tight correlation of the index maximum vs.\ cluster [Fe/H] 
allover the entire metallicity range for stellar population
with [Fe/H]~$\gtrsim -2.0$. Relying on a theoretical calibration 
of the feature, we trust to effectively infer cluster metallicity within a 
typical uncertainty of 0.1--0.2~dex, depending on RGB luminosity sampling 
of the observations.

A handful of stars (5--10 objects) is required for the method to be applied, 
with low-metallicity stellar populations more easily managed, being 
Fe5270$_{\rm max}$ located within the few brightest RGB stars of the system. 
In any case, we show that even the observation of a coarse stellar set would 
allow us to place a confident lower limit to cluster metallicity.
\end{abstract}

\keywords{stars: abundances --- globular clusters: general --- 
open clusters and associations: general --- Galaxy: disk}

\section{Introduction}
A fair estimate of metal abundance in stellar systems still remains a 
central issue for any detailed assessment of the other cluster distinctive properties.
Metallicity modulates, in fact, both effective temperature and apparent 
colors of stars, while the internal composition affects the nuclear engine,
and therefore stellar lifetime.
Even in case of resolved stellar systems, this entangled behavior may lead to 
a biased interpretation of cluster age, based on the CMD morphology, an effect 
often referred to as the ``age-metallicity dilemma'' \citep{rb86,worthey94}.
Such induced age uncertainty also reflects in the distance determination of 
star clusters, as far as one tries to compare the apparent magnitude of the Main 
Sequence Turn Off (TO) point with the appropriate theoretical luminosity to derive 
therefrom the distance modulus.
One further difficulty also deals with the proper assessment of dust reddening, that 
may affect CMD morphology leading, in general, to an artificially enhanced value 
of [Fe/H].\footnote{In terms of color excess, this effect can be quantified
in $E(B-V) \simeq 0.14\,\Delta {\rm [Fe/H]}$ \citep{buzzoni95a}.}  

To overcome these problems, one would like to preliminarily derive 
the cluster metallicity from accurate abundance analysis of individual stars
(typically red giants, due to their brighter intrinsic luminosity at optical
wavelengths) through high-resolution spectroscopy \citep[e.g.,][]{carretta97,kraft03}.
This delicate task, however, is extremely
time consuming, and a far more straight shortcut is often pursued relying on integrated 
cluster spectroscopy, usually taken at much lower resolution. Narrow-band 
spectrophotometric indices, as derived from integrated 
low-res ($\sim$~6-8~\AA\ FWHM) observations, usually provide the basic diagnostic 
scheme to match theoretical models from stellar population synthesis and
derive therefrom cluster properties \citep[e.g.,][]{pacheco98,beasley02,strader04}. 
The Lick system \citep{burstein84,wortheyetal94,trager98} 
stands out as the widest and most popular reference, assuring  
a systematic coverage of the main spectral features across the 
4100--6400~\AA\ wavelength range, easily accessible from ground-based observations.

Besides the advantage of this strategy, one can still
question whether it effectively allows us to disentangle any age-metallicity 
degeneracy. In fact, integrated indices are sensitive both to intrinsic 
elemental abundance and to the temperature distribution of the underlying 
stellar population, thus delivering a composite and likely non-univocal piece of 
information \citep[see a discussion in][]{buzzoni95b,worthey95,
tantalo04}.

An alternative, and possibly more proficient way out, at least for resolved
stellar clusters, can be envisaged and will be explored in this paper. 
It relies on a minimal ``tuned'' sampling of individual red giant 
stars of a cluster, observed at low spectral resolution such as to derive 
the maximum strength for the Fe5270 index, which is by far the most 
popular one among the several Fe~{\sc i} features included in the Lick system.
As we will see in Sect.~2, for their special physical properties,
Fe~{\sc i} features prove to be ideal candidates for a reliable 
``snapshot'' estimate of stellar system metallicity.

Our method complements previous studies in the same line. 
For example, by relying on low-resolution near-IR observations of the strong 
Ca~{\sc ii} triplet among RGB stars, \citet{dacosta98} and \citet{parisi09} 
recently recovered the [Fe/H] value for the SMC and its cluster system
within less than a 0.2~dex uncertainty compared, for instance, to the 
high-res analysis of \citet{battaglia08}.

In the following discussion, we will outline our method in Sect.~3, and 
propose a plain theoretical calibration based on population synthesis 
models. Theoretical predictions will be compared with a grid of Galactic 
globular and open clusters spanning the whole range of metallicity.
In Sect.~4 we will summarize our conclusions and also briefly outline 
further applications of the method, also in view of the deeper forthcoming
surveys of external stellar systems (i.e., extragalactic star clusters
around and within nearby galaxies in the Local Group, etc.) 
taken with the new-generation telescopes.

\section{Narrow-band indices of stars}

Extended grids of synthetic stellar spectra, from which to derive Lick-index 
behavior vs.\ $\log T_{\rm eff}$, $\log{g}$, and [Fe/H], have been computed, 
among others, by \citet{mould78}, \citet{barbuy94}, \citet{tripicco95},
\citet{chavez96}, \citet{barbuy03}, and \citet{bertone02}.
In addition,  semi-empirical calibrations based on more or less complete 
samples of disk stars are due, among others, to \citet{buzzoni92}, 
\citet{buzzoni94}, \citet{gorgas93}, \citet{wortheyetal94}, \citet{borges95} 
and \citet{franchini04}. The latter results are often summarized in 
analytical sets of so-called ``fitting functions''.

Both approaches resulted in a complementary contribution as far as the physical behavior 
of the different spectral features vs.\ stellar fundamental parameters is
concerned. In this respect, a recognized advantage of synthetic spectra
is to directly explore index changes within a ``controlled'' grid of input parameters. 
However, models are still largely inadequate in their physical treatment 
of cool stars, 
\citep[i.e., $T_{\rm eff} \lesssim 4000$~K, including late-K and M spectral types; see][]{bertone04,bertone08}, when convection and the increasing effect of molecular 
opacity severely perturb the atmosphere structure and the emerging spectral energy 
distribution of stars at optical wavelength.

From its side, the empirical fitting-function technique has the advantage
of giving a simple and accurate summary of index behavior across
the full parameter domain of real stars. This partly overcomes any modeling
uncertainty, and the analytical set of equations can easily be implemented in 
theoretical models of stellar populations to compute the integrated indices
of a system without requiring any spectral input for individual stars
\citep{buzzoni92,buzzoni94,worthey94,maraston01,maraston05}. 
As a drawback of this approach, however, one has to remark the evident
limitation of any empirical star sample (forcedly dominated by a 
component of solar neighbors) to properly probe the full metallicity scale
outside the solar range.

\subsection {Fe5270 diagnostic properties}

Among the set of Lick indices in the optical wavelength range,
the Fe~{\sc i} lines reveal potentially the most promising ones for the 
present analysis. This is especially true for the Fe5270 feature, 
that begins to appear in stars cooler than $T_{\rm eff}=7500$~K and reaches a maximum 
equivalent width about 4200--4600~K \citep{buzzoni94}.
The latter is the typical temperature of faint low-main sequence dwarfs 
and, most importantly, also of bright MK~III giants, which populate the 
AGB/RGB region of the CMD diagram.

Although not so prominent, the Fe5270 feature (and its close neighbor
Fe5335) can easily be recognized also in the integrated spectra of elliptical 
galaxies \citep{worthey92,trager98}, and its presence is known
to trace the relative contribution of low-RGB stars in the 
galaxy stellar population \cite{buzzoni95a}.

\begin{figure}
\centering
\includegraphics[width=\hsize]{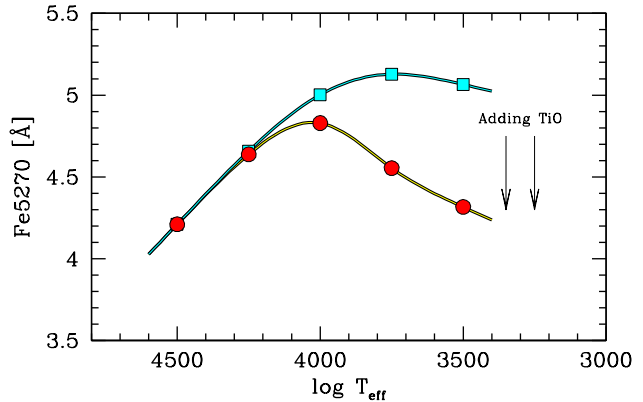}
\caption{The Fe5270 index as a function of \teff\ from {\sc Bluered} 
synthetic stellar spectra of model atmospheres with $\log g = 3.0$~dex 
and solar metallicity (dots). Square markers trace the raw Fe5270 index trend
by neglecting the TiO molecular opacity.}
\label{fig:index_tio}
\end{figure}

In Fig.~\ref{fig:index_tio} we explored in finer detail the Fe5270 diagnostic 
properties from the theoretical point of view, by means of an {\it ad hoc} grid 
of {\sc Bluered} synthetic spectra \citep{bertone02,bertone08}. Spectral synthesis
relied on {\sc Synthe} code \citep{kurucz93,sbordone04} and the grid
of {\sc Atlas9} revised model atmospheres \citep{castelli03} along the 
3500--4500~K \teff\ range.
For $T_{\mathrm{eff}} \gtrsim 4000$~K, the main contributor to line blanketing,
along the three wavelength bands that define the index, is found to be
Fe~\textsc{i}, with a fine structure of more than 200 lines; particularly 
strong is, of course, the $\lambda$5269.5 line.
However, in this temperature regime, also important is the contribution of
neutral Titanium and Chromium and, secondly, Calcium and Cobalt.

The TiO opacity becomes increasingly stronger with decreasing temperature, and
becomes dominant for $T_{\mathrm{eff}} \lesssim 3750$~K. As a result, 
the prevailing effect of the TiO molecular bands on the Fe5270 index is
in the sense of depressing the red-side pseudo-continuum, thus reducing 
the apparent strength of the index. According to the models, one may even
conclude that the TiO molecule is actually the main responsible for 
constraining in temperature the Fe5270 peak value (see, again, 
Fig.~\ref{fig:index_tio}).

We can probe index sensitivity also by means of empirical data.
From the \citet{buzzoni94} fitting function we can write 
in fact:
\begin{equation}
{\rm Fe5270}_{\rm max} = 1.15~{\rm [Fe/H]} +0.25 \log g +3.80,
\label{eq:fe52}
\end{equation}
that fairly well compares with the {\sc Bluered} theoretical 
prediction,\footnote{The [Fe/H] dependence of the {\sc Bluered} synthetic index 
has perhaps to be considered as an upper limit, as molecular absorption in the 
relevant wavelength range should likely be stronger than predicted by models at solar and 
super-solar metallicity. According to our experiments, this makes the current 
{\sc Bluered} output to overestimate by $\sim$0.5~\AA\ the real 
Fe5270 index strength as observed for metal-rich stars.} namely
\begin{equation}
{\rm Fe5270}_{\rm max} = (1.68-0.1 \log g)~{\rm [Fe/H]} +4.80.
\label{eq:0}
\end{equation}
According to \citet{wortheyetal94} models we have
\begin{equation}
{\rm Fe5270}_{\rm max} = 1.29~{\rm [Fe/H]} +(0.11 \log g -0.51) \log g +4.54,
\label{eq:fe52w}
\end{equation}
while \citet{gorgas93} results provide\footnote{Note that the \citet{wortheyetal94} 
and \citet{gorgas93} results are not completely independent as both rely
on a largely overlapping stellar sample. One relevant difference,
however, is that \citet{gorgas93} use the $V-K$ color instead of 
$T_{\rm eff}$ for their fitting functions.}
\begin{eqnarray}
{\rm Fe5270}_{\rm max}  = & (1.79+0.05 \log g +0.19~{\rm [Fe/H]})~{\rm [Fe/H]} + \nonumber \\
                          & +(0.13 \log g -0.31) \log g +4.55.
\label{eq:fe52g}
\end{eqnarray}
Note from the equations the weak index sensitivity to stellar
surface gravity. For example, any change of one dex in $\log g$ 
only contributes at most by $\pm 0.3$~\AA\ to the Fe5270$_{\rm max}$
variation. In terms of overall properties of a stellar aggregate, 
this figure reflects in a change of TO stellar mass of roughly 0.3 dex, which
implies a variation of nearly one order of magnitude in stellar lifetime. 
Therefore, as a major conclusion, we have that Fe5270$_{\rm max}$ in a stellar 
population {\it is nearly independent from age}.

\begin{figure}
\psfig{file=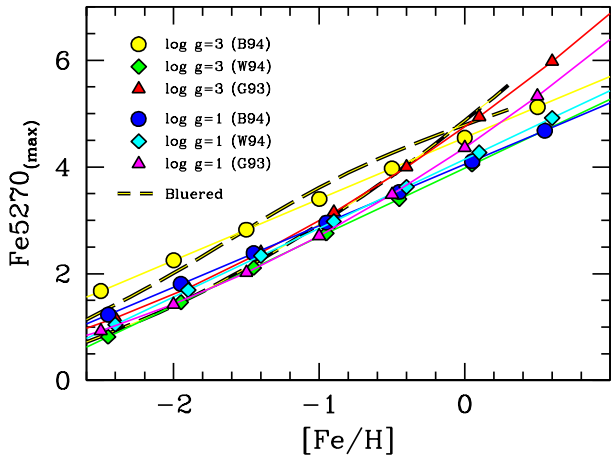,width=\hsize,clip=}
\caption{Fitting-function predictions for Fe5270$_{\rm max}$ vs.\ [Fe/H] 
relationship according to \citet{buzzoni94} (B94, see eq.\ref{eq:fe52}), 
\citet{wortheyetal94} (W94, eq.\ref{eq:fe52w}) and \citet{gorgas93}
(G93, eq.\ref{eq:fe52g}).
Two relevant cases for stellar gravity are considered, namely $\log g = 3$ and 1 dex.
The theoretical values from {\sc Bluered} synthetic spectra are also 
overplotted (dashed line).
}
\label{f3}
\end{figure}

A comparison of the different calibrations is shown in Fig.~\ref{f3},
where we report the output index for two reference values of stellar 
gravity (namely, $\log g = 1$ and 3 dex) appropriate for red giants.
Within a remarkably good agreement among the different datasets, 
compared to \citet{buzzoni94} calibration, one notes from the plot that 
\citet{wortheyetal94} fit displays an even lower dependence on $\log g$, while 
\citet{gorgas93} fit stands out for its steeper trend with metallicity, 
especially at super-solar regimes. Overall, we can conclude that the {\it direct} 
Fe5270 dependence on Fe abundance turns out to be $\Delta {\rm Fe5270} =
\alpha \Delta {\rm [Fe/H]}$, 
with $\alpha = 1.2\rightarrow 1.4$.

\begin{figure}
\centerline{
\psfig{file=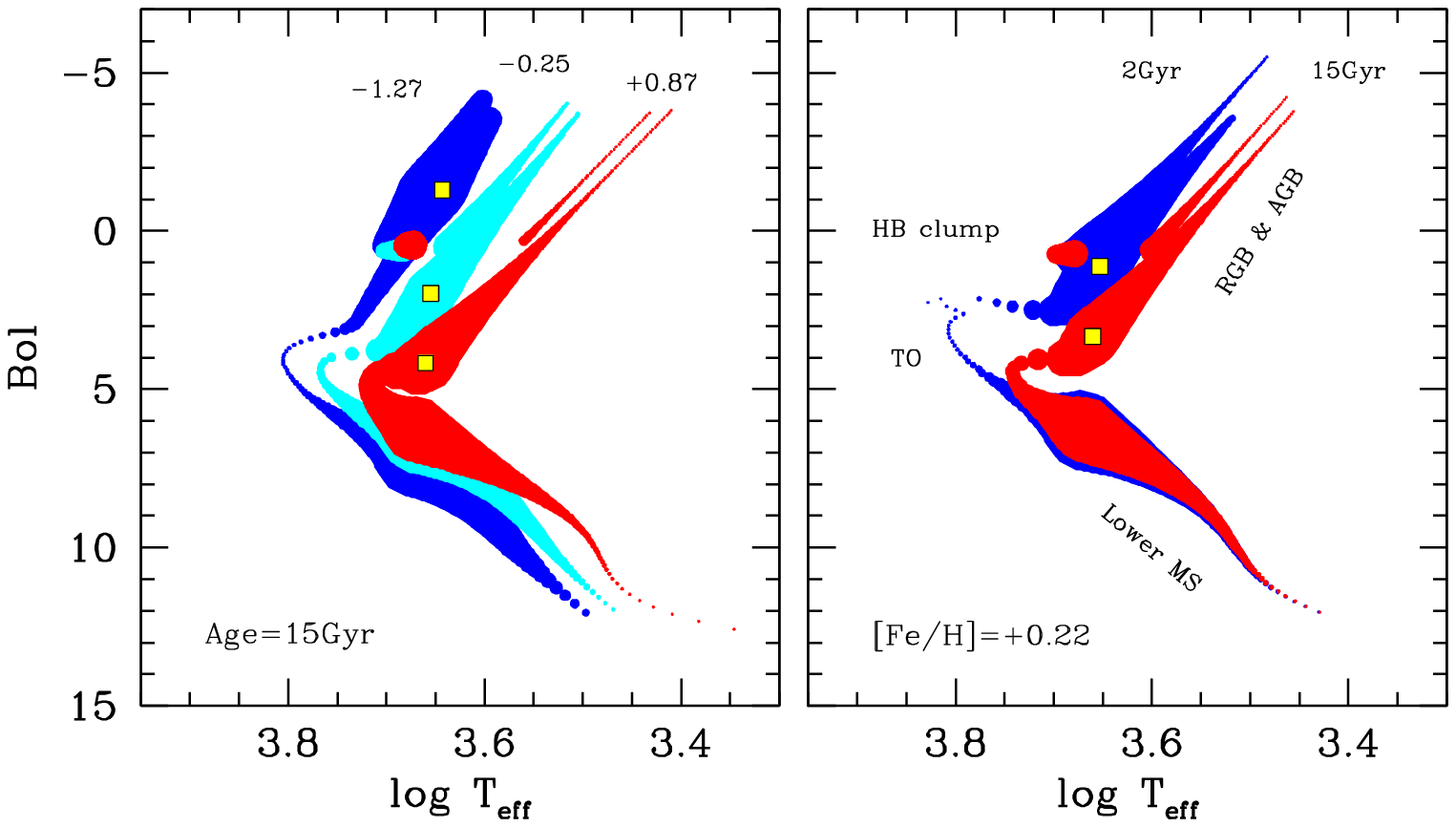,width=\hsize,clip=}
}
\caption{The expected Fe5270 Lick index strength in SSPs for an illustrative set of 
theoretical CMDs,
from \citet{buzzoni89,buzzoni95a} models. Left panel explores
the index change vs.\ [Fe/H] (as labeled on the plots) for a set of 15 Gyr 
populations, while right panel takes into account the index variation for fixed values 
of metallicity with changing SSP age (from 15 and 2 Gyr). The width of the
isochrones is proportional to the local index value. Small square markers locate the
maximum value, throughout.}
\label{f2}
\end{figure}

\section {Tracing stellar populations}

In addition to its direct sensitivity to [Fe/H], a major supplementary advantage 
of using Fe5270 feature as metallicity tracer in stellar populations resides in the 
special property of the index to always peak {\it within} the temperature 
range sampled by red giants in stellar systems richer than 
[Fe/H]~$\gtrsim -2$~dex.\footnote{Apart from Fe5270, two other Lick indices display 
a similar behavior: the G-band and the CN molecular indices 
\citep[see][]{gorgas93}.
For its atomic origin, however, Fe5270 should in principle be a more accurate and 
unbiased metallicity tracer.}

Conversely, this is not the case, for instance, of other stronger features, used
to derive popular indices like the Lick Mg$_2$ or the near-IR CaT triplet 
index \citep{jones84,idiart97,cenarro01}. Both these features peak, in fact, 
among giants and dwarfs at the coolest temperature tail of each CMDs, and their dependence 
on SSP metallicity is, therefore, naturally prone to more entangled age effects and to other 
model-dependent features, such as convection and mass loss properties (both 
constraining the AGB/RGB tip location).

The Fe5270$_{\rm max}$ location for the illustrative case of different simple stellar 
populations (SSPs), with varying either age or [Fe/H], is shown in 
Fig.~\ref{f2}. A glance to the figure makes clear that Fe5270 homogeneously 
probes the ubiquitous component of  K1-K3~III stars within each stellar system.
 
The collection of empirical estimates of Fe5270$_{\rm max}$ for several open and globular 
clusters of the Galaxy, according to the \citet{gorgas93} database, is summarized in 
Fig.~\ref{f4}. These results are collected in Table~1, where we report for each 
cluster the strongest value reached by the Fe5270 index among the sampled stars, together with 
the sample size in each system, and the adopted cluster metallicity.
Note that, as Fe5270 strength peaks among increasingly brighter red giants with decreasing
metallicity (see left panel of Fig.~\ref{f2}), just a handful set of bright 
stars in metal-poor globular clusters is sufficient to suitably pick up Fe5270$_{\rm max}$.
A slightly deeper spectroscopy is required, instead, for metal-rich
systems, but even in this case a dozen of stars taken at low spectral resolution 
are fully sufficient to suitably constrain the index tip.
As an interesting example, in this regard, we also overplot in Fig.~\ref{f4}
the full sample of field giants (class MK III or $1 \le \log g \le 3$) studied by 
\citet{buzzoni94}, and \citet{buzzoni01}; it is evident
that the upper envelope of the Fe5270 distribution 
for the entire stellar dataset effectively matches the overall cluster trend. 

The data of Fig.~\ref{f4} are also compared with theoretical predictions from population 
synthesis models. The displayed models refer to 12.5~Gyr SSPs with different metallicity 
from \citet{buzzoni89,buzzoni95a}. The case of younger (5~Gyr) populations is also accounted for
the metallicity range of open clusters.\footnote{For young ($t \lesssim 6$~Gyr) and 
extremely metal-poor ([Fe/H]~$\lesssim -2$) SSPs, however, models indicate that
red giants are ``too blue'' (i.e., too warm) for the Fe5270 feature to span its
full range and reach its maximum. In these case the Fe5270 index mat actually drop to 
nominal values and even remains undetected in the spectra of these stars.}

\begin{deluxetable}{lcccr}
\tabletypesize{\scriptsize}
\tablecaption{Summary of relevant data for Galactic open and globular clusters$^{(a)}$}
\tablewidth{0pt}
\tablehead{
\colhead{Name} & \colhead{Type$^{(b)}$}  & \colhead{[Fe/H]} & \colhead{Fe5270$_{\rm max}$} & \multicolumn{1}{c}{No. of}  \\ 
\colhead{ }           & \colhead{ } &   & \colhead{[\AA]}         & \multicolumn{1}{c}{sampled stars} \\
}
\startdata
NGC~188     & O    & --0.02  &  4.03  & 24$\qquad$ \\
NGC~7789    & O    & --0.08  &  4.45  & 12$\qquad$ \\
M67         & O    & --0.10  &  3.89  & 20$\qquad$ \\
M71$^{(c)}$ & G    & --0.73  &  3.18  & 36$\qquad$ \\		     
M5          & G    & --1.27  &  2.94  &  5$\qquad$ \\
M10         & G    & --1.52  &  2.32  &  4$\qquad$ \\
M13         & G    & --1.54  &  2.09  &  3$\qquad$ \\
M3          & G    & --1.57  &  2.57  &  3$\qquad$ \\
M92         & G    & --2.28  &  1.43  &  4$\qquad$ \\
\enddata
\tablenotetext{a}{From \citet{gorgas93} Table~3. Metallicity of globular clusters is from Harris(1996),
while for open clusters it derives from the WEBDA database \citep{paunzen08};}
\tablenotetext{b}{O = Open cluster; G = Globular cluster;}
\tablenotetext{c}{A supplementary entry exists for this cluster with Fe5270$_{\rm max} = 3.99$ but
star has been dropped by \citet{gorgas93} as possible field interloper.}
\label{tab1}
\end{deluxetable}

\section{Results and conclusions}

\begin{figure}
\centerline{
\psfig{file=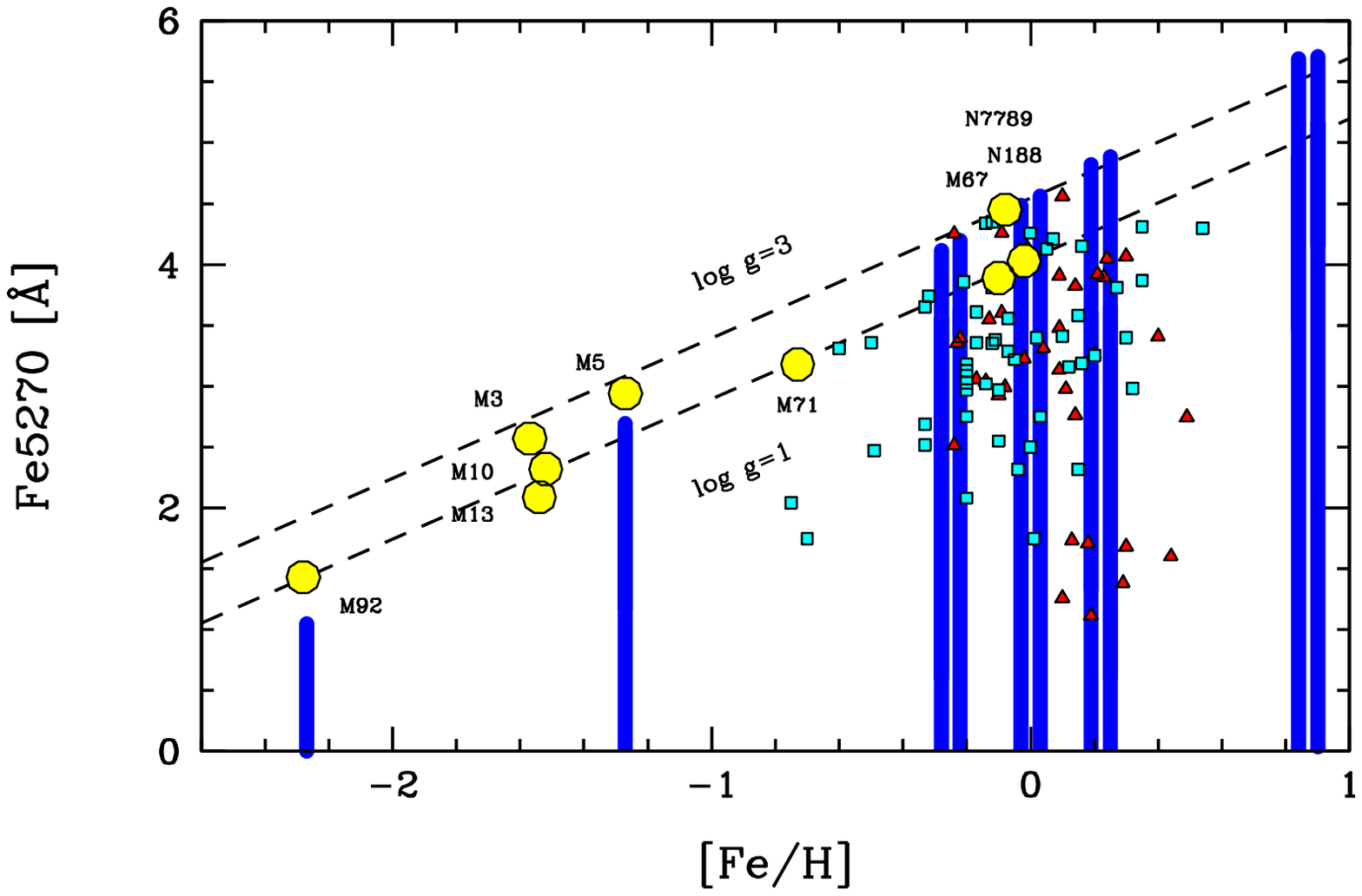,width=\hsize,clip=}
}
\caption{The observed distribution of Fe5270$_{\rm max}$  from the
star cluster sample of \citet{gorgas93}.
Big dots mark the strongest value of Fe5270 among the sampled giant stars in each
open and globular cluster, as summarized in Table~1. 
Only {\it bona fide} cluster members are considered, according to \citet{gorgas93}
classification. The two field giants samples studied by \cite{buzzoni94,buzzoni01}
(small triangles and squares, respectively) are also overplotted, considering only giant
stars (i.e., $1 \le \log{g} \le 3$) with confident fundamental parameters. 
Observations are compared with stellar fitting functions for fixed gravity,
i.e.\ $\log{g} = 3$ and 1 as labeled, according to eq.(\ref{eq:fe52}).
Also theoretical predictions from stellar population synthesis are reported, 
according to \citet{buzzoni89,buzzoni95a} SSP models (thick vertical bars);
an age $t = 12.5$~Gyr is adopted throughout, with the additional
cases of 5~Gyr populations for the metal-rich ([Fe/H]~$\ge -0.5$) regime
more pertinent to Galactic open clusters (left ``column'' at each relevant metallicity).
}
\label{f4}
\end{figure}

A fit to all synthesis models of Fig.~\ref{f4}, excluding the [Fe/H]~$= -2.27$ case, provides
\begin{equation}
{\rm Fe5270}_{\rm max} = 1.40_{(\pm 0.03)} {\rm [Fe/H]} +4.52_{(\pm 0.02)},
\label{eq:fitssp}
\end{equation}
with the displayed coefficient uncertainty at 1$\sigma$ level.
Compared to the original stellar fitting function of eq.~(\ref{eq:fe52}),
we report here a slightly steeper slope (i.e., a more sensitive dependence) in the 
Fe5270 vs.\ [Fe/H] SSP relationship (namely $\alpha = 1.40$ here vs.\ 1.15 of eq.~\ref{eq:fe52}).
This is due to the reinforcing effect of the $\log{g}$ variation, which increases among
metal-rich giants, for fixed SSP age.

Entering eq.~(\ref{eq:fitssp}) with the cluster values for Fe5270$_{\rm max}$ from Table~\ref{tab1}
we could assess the statistical performance of our theoretical calibration as [Fe/H] predictor.
One has to bear in mind, however, that this procedure is clearly biased toward underestimating
cluster metallicity as we might be missing the real Fe5270 maximum.
Of better interest may rather be to study the distribution of positive and negative residuals
(in the sense ``observed - predicted'' [Fe/H]), separately. In summary, the relevant figures 
are the following:
\begin{equation}
\sigma_{\rm [Fe/H]}^{\rm tot} = \pm 0.21~{\rm dex}
\left\{
\begin{array}{ll}
\sigma^+_{\rm [Fe/H]} = 0.13~{\rm dex}\\
    \\
\sigma^-_{\rm [Fe/H]} = 0.26~{\rm dex}
\end{array}
\right.
\end{equation}
where the rms for the total sample residuals (l.h.\ side of the equation) comes in fact
from a skewed distribution of positive ($\sigma^+_{\rm[Fe/H]}$) and negative 
($\sigma^-_{\rm [Fe/H]}$) residuals.
For the previous arguments, the $\sigma^+_{\rm [Fe/H]}$ value may actually provide a 
more genuine estimate of the real method performance, once a fair spectroscopic sampling 
of cluster stellar population can be assured.
 
In conclusion, the claimed advantage of our diagnostic technique for a ``snapshot'' estimate 
of metal abundance in stellar systems (providing one resolves at least the brightest part
of the CMD) can be briefly summarized as follows:

1) As Fe5270$_{\rm max}$ is negligibly affected by stellar gravity, its value 
within a stellar population is virtually age independent, and may therefore provide
an unbiased measure of metallicity within a 0.1--0.2~dex internal uncertainty.
The Fe5270$_{\rm max}$ vs.\ [Fe/H] correlation is expected to strictly hold for every 
stellar population with [Fe/H]~$\gtrsim -2.0$. 

2) As far as we know, within this framework, the Fe5270 index performance is definitely 
better than that of any other stronger feature. For instance, by relying on Mg$_2$ or 
CaT$^*$ indices \citep[see][]{cenarro01}, our procedure delivers
cluster metallicity for the sample of Table~\ref{tab1} 
with a minimum\footnote{Because one should also account for an additional uncertainty in case
of unknown cluster age)} internal accuracy of $\sigma({\rm [Fe/H]})_{\rm Mg2} \sim 0.25$~dex 
and $\sigma({\rm [Fe/H]})_{\rm CaT^*} \sim 0.42$~dex, respectively.
   
3) Compared to other spectroscopic methods, ained at deriving cluster
metallicity from low-resolution spectra of their bright stars, our method
has the superior advantage that it does not require any {\it a priori} assumption/inference
about stellar temperature and gravity. It is distance and reddening independent, 
and uses an Iron-dominated feature (instead of any $\alpha$-element 
feature, as for Mg and Ca indices) to infer [Fe/H].
   
4) Low-resolution spectroscopy ($R = 600 \rightarrow 800$) is required for
the method, planning just a ``one-shot'' observation of the brightest stars of a given 
cluster with a standard multi-object (MOS) instrumental setup. Compared to broad-band 
imaging, one could envisage to derive useful spectroscopic information for stars 
about 3--4 mag brighter than the imaging limit.

5) As an added value, in addition to the Fe5270 index, spectra would naturally 
deliver supplementary information on other popular Lick indices, like Mg$_2$,
H$\beta$, etc., at close wavelength.

6) A minimal set of giant stars is required (5--10 objects), hopefully spanning the whole 
luminosity range. 
Low-metallicity stellar populations would be the favorite ones, 
as Fe5270$_{\rm max}$ could easily be located among the very few outstanding 
stars of the cluster. Deeper observations (down to the base of the RGB) 
would be required, instead, for metal-rich populations. 
According to \citet{rb86}, however, the  luminosity-specific full
number of Post-MS stars in a SSP with TO mass, M$_{\rm TO}$, is
\begin{equation}
{{n}\over{L}} \simeq 0.03\,M_{\rm TO}^{-2.72}.
\end{equation}
This means, for instance, that by sampling some $10^3\,{\rm L}_\odot$ in the uncrowded
external region of a globular cluster one may successfully pick up Fe5270$_{\rm max}$ among the few dozen brightest stars in the field.
In any case, the observation of just the few outstanding members would confidently place
a lower limit to cluster metallicity in lack of any other piece of information.

\acknowledgments
It is a pleasure to thank Guy Worthey, the referee of this paper, for his valuable 
suggestions and for an accurate review of our work.
We also acknowledge partial financial support from Mexican CONACyT under 
grant 49231-E.

\clearpage
\end{document}